\documentclass{article}
\usepackage{amsmath,epsfig,times,url}

\title{Simulating the Electroweak Phase Transition:
 Sonification of Bubble Nucleation}

\usepackage{graphicx} 
\usepackage{epstopdf} 

\author{R. Michael Winters\footnote{MikeWinters10@gmail.com},~ Andrew Blaikie\footnote{ABlaikie13@wooster.edu}\\Department of Physics, College of Wooster  \and Deva O'Neil\footnote{DONeil@bridgewater.edu}\\Department of Physics,  Bridgewater College } 

\date{}
\begin{document}
\maketitle

\begin{sloppy}

\begin{abstract}

As an applicaton of sonification, a simulation of the early universe was developed to portray a phase transition that occurred shortly after the Big Bang.  The Standard Model of particle physics postulates that a hypothetical particle, the Higgs boson, is responsible for the breaking of the symmetry between the electromagnetic force and the weak force.  This phase transition may have been responsible for triggering Baryogenesis, the generation of an abundance of matter over anti-matter.  This hypothesis is known as Electroweak Baryogenesis.  In this simulation, aspects of bubble nucleation in Standard Model Electroweak Baryogenesis were examined and modeled using Mathematica, and sonified using SuperCollider3.  The resulting simulation, which has been used for pedagogical purposes by one of the authors, suggests interesting possibilities for the integration of science and aesthetics as well as auditory perception. The sonification component in particular also had the unexpected benefit of being useful in debugging the Mathematica code.  \end{abstract}

\maketitle

\section{Introduction and Motivation}

\subsection{Motivation for Sonification}

Questions regarding theoretical particle physics and cosmology are of increasing interest to the scientific community due to the current operation of the Large Hadron Collider.  On the experimental end, sonification has been explored as a tool for particle detection \cite{Vogt:2010}.  On the theoretical end, sonification has a history of application through simulations of rigorous and complex theoretical models \cite{Vogt:2010Diss}\cite{Vogt:2010Spin}.  Perhaps the most important benefit of sonification to the scientific community is an informed and interested public.  Sonifications, visualizations, and simulations provide a point for public access to the otherwise challenging aspects of contemporary physics.  Additionally, the stochastic and probabilistic nature of this process lends itself to discussions of aesthetics in the sciences \cite{Voss:1978}.

This simulation was developed to portray the universe as it underwent a fundamental change to a lower-energy state about $10^{-10}$ seconds after the Big Bang.  This ``phase transition" lasted $10^{-13}$ of a second and left the universe with a permanent abundance of matter over anti-matter.   The simulation illustrates this very brief but important point in the evolution of the universe.  

In this simulation, the phase transition has been portrayed as proceeding by bubble nucleation.  Similar phase transitions are familiar from everyday life; as water vapor cools in a cloud, the water condenses into its liquid phase.  Impurities or dust particles in the cloud serve as centers for the nucleation of bubbles, and the bubbles expand as the surrounding air molecules are tipped into the lower energy state, triggering a first-order phase transition from liquid to solid. 

For the purposes of this research, we assume that a similar first-order transition took place in the early universe.  Our final product is a set of 13-second movie files (.mov) showing circular bubbles (representing areas of broken electroweak symmetry, the lower-energy state) appearing in a background of static (representing energy fluctuations in a hot universe).  The timing and location of the bubble formation is governed by a probability function rather than hard-coded, so that each time the code is run a different distribution of bubbles result.  The formation of each bubble is accompanied by a ``chime" carrying audio information about the position of the bubble.  As the bubbles expand and fill the screen, there is an increase in volume and change in timbre to reflect that more and more of the universe has transitioned to the lower-energy state, producing the abundance of matter over antimatter.

\subsection{Conditions for Baryogenesis}

Although antimatter can be created in small quantities in particle accelerators, by current observations the visible universe is practically void of anti-matter on all scales.  Most of the mass in the universe is baryonic.  A baryon, of which protons and neutrons are the most common example, is a particle made up of 3 quarks.  Anti-baryons are composed of 3 anti-quarks.  The asymmetry between the number of baryons per unit volume ($n_{b}$ ) and the number of anti-baryons per volume ( $\bar{n}_{b}$ ) is defined as

\begin{equation} \frac{n_b - \bar{n}_{b}}{n_b + \bar{n}_{b}}.\end{equation}

In practice, baryons will annihilate with anti-baryons until the supply of anti-baryons is exhausted, producing photons:

\begin{equation}B + \bar{B} \rightarrow \mbox{photons}.\label{pho}\end{equation}
The baryon asymmetry is given experimentally by the baryon-to-photon ratio:

\begin{equation} \eta \equiv \frac{n_b}{n_\gamma} = 6\cdot 10^{-10},\end{equation}
where $n_\gamma$ is the number density of cosmic microwave background photons, assumed to have been produced by the annihilation process in eq. \ref{pho}. Thus, for every billion baryons originally in the universe, there must have been a deficit of about one anti-baryon.  This matter-antimatter asymmetry remains one of the most important unanswered questions in cosmology. 

Historically, the abundance of matter was attributed to the initial conditions of the Big Bang. In 1967,  Andrei Sakharov~\cite{Sakharov:1967dj} proposed three conditions that would allow for Baryogenesis, the production of more matter than anti-matter:
\begin{eqnarray}\label{csvA}\nonumber
&&\text{A) baryon number violation in the laws of nature,} \\\nonumber
&&\text{B) CP-Violation, and} \\
&&\text{C) a departure from thermal equilibrium.} 
\end{eqnarray}   


Baryon number is defined as the number of baryons minus anti-baryons in the universe.  Sakharov's first condition requires that there be some process that can generate an excess of baryons. CP-Violation, or charge-parity violation, occurs when left-handed particles and right-handed antiparticles do not behave in the same way.  If there is not enough CP-Violation, the anti-matter process that generates an excess of anti-baryons would occur at the same rate as the process that generates an excess of baryons; thus no imbalance could be formed. Finally, if the universe were in thermal equilibrium, the reverse reactions would occur at the same rate. Hence, a departure from thermal equilibrium is necessary to achieve the matter anti-matter imbalance. All three of these conditions may be applicable to the Electroweak Phase Transition, leading to speculation that the breaking of the electroweak symmetry may have been the source of the matter-antimatter asymmetry.

\subsection{The Electroweak Phase Transition \& Baryogenesis}

When the universe was very young, it was highly energetic.  At these high energies, the electromagnetic and the weak nuclear force were symmetric and mediated by four massless gauge bosons. The Standard Model of particle physics hypothesizes the existence of an additional boson (the Higgs particle); as the universe cooled, the Higgs field would have broken the electroweak symmetry as it transitioned to its vacuum state, provided that its vacuum energy was non-zero.  (The actual vacuum expectation value that fits current data is 246 GeV.)  This breaking of electroweak symmetry left the photon massless and gave mass to the $W^\pm$  and $Z$ bosons of the weak force.  This moment is of particular interest to Baryogenesis. The Higgs field does not acquire its vacuum energy uniformly throughout space; in the scenario of electroweak baryogenesis, regions of space where the Higgs field transitions to its vacuum state form bubbles of broken symmetry, which would expand or shrink depending on their radii. This process is called bubble nucleation and it is within the bubble ``walls" that Baryogenesis could occur.  

All three of Sakharov's condtions may be present in the Electroweak Phase Transition.  Baryon number violation occurs in standard electroweak theory; although it is suppressed at low temperatures, baryon number violation becomes stronger at high temperatures \cite{Dine:1989kt}.  The following toy example illustrates how Electroweak Baryogenesis could proceed:  Suppose the process to generate an excess of baryons is $X\to Y+B$. If there is enough CP-Violation in the model then this process may occur more often than its anti-matter equivalent process $\bar{X}\to \bar{Y}+\bar{B}$. If the phase transition is significantly first order, or extremely abrupt, the region of rapidly changing Higgs field would be out of thermal equilibrium. Hence the reverse processes $\bar{Y}+\bar{B}\to \bar{X}$ and $Y+B\to X$ would not occur at the same rate as the forward processes, perpetuating the imbalance. The excess of baryons would be swept into the bubble and acquire mass. Since the new conditions are drastically different inside the bubble, the excess matter would persist and baryon number would be violated.  Theoretical studies of baryon number violating processes have shown that such a process is possible within the walls \cite{Trodden:1999}.

In this simulation, we assume, for simplicity, that the breaking of electroweak symmetry is governed by the Higgs content of the Standard Model (a single Higgs particle).  This is not the most realistic electroweak baryogenesis scenario; for one thing, the Standard Model does not provide enough theoretical CP-Violation for Electroweak Baryogenesis to generate the observed excess of matter.  Furthermore, the mass of the Higgs boson would have to be less than roughly 40 GeV to create a sufficiently large departure from thermal equilibrium \cite{Dine:1992ad}. This mass is improbable because particle accelerators would have already discovered a Higgs particle this light; an electroweak phase transition in line with current experimental bounds on the Higgs mass would be at best only weakly first order.  Other models such as Supersymmetry or a Two-Higgs Doublet Model may allow for more CP-Violation and possibly a higher Higgs mass. Thus, Electroweak Baryogenesis is not ruled out in those models.

This paper provides a model of bubble nucleation in Electroweak Baryogenesis assuming a Standard Model Higgs sector and a Higgs mass of 35 GeV.  Although this mass is unrealistic according to current experimental bounds, it is a useful starting point for modeling Electroweak Baryogenesis. The mathematical formalism will be simplified in favor of a more rigorous explanation of the sonification techniques.  For the full mathematical formalism see ref. \cite{Blaike:2011}.

\section{Effective Potential, Critical Radius, \& Wall Velocity}	

The evolution of the Higgs field with time (or equivalently, temperature) is governed by its effective potential (see Fig. \ref{EffectivePotenLow}).  The equations used to define the effective potential in this work were taken from \cite{Dine:1992ad}. To create the simulation, one must derive values for the critical radius and wall velocity of the bubbles.  For a Higgs mass of 35 GeV, the critical temperature was found to be $T_c=71.4$ GeV.   This is the temperature at which a second minimum forms in the potential.  Below this temperature, the state of broken electroweak symmetry (a non-zero vacuum expectation value for the Higgs field) is energetically favorable.  

\begin{figure}[ht] 
	\centering
	\includegraphics[width=1\linewidth]{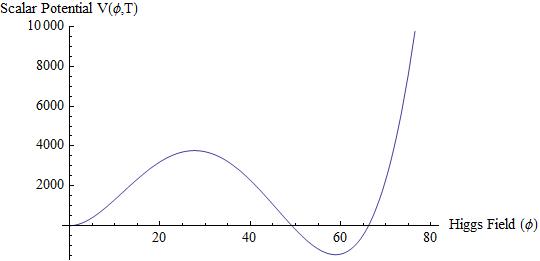} 
	\caption{The effective potential for a Higgs mass of 35 GeV at 71.4 GeV temperature. As the universe cools to the ``critical temperature" $T_c$, the Higgs potential forms a second minimum.  The deepening of this minimum (shown here at $60 GeV$) as the cooling continues forms a ``true vacuum" (global minimum).  Energy fluctuations allow the Higgs field to transition into the true vacuum, forming a bubble of broken electroweak symmetry.}
	\label{EffectivePotenLow}
\end{figure}

Based on this effective potential, the critical radius was derived. The critical radius (which depends on temperature) is the minimum size of bubble that will allow the bubble to survive and grow; bubbles that are nucleated with subcritical radii do not gain enough energy to overcome surface tension, and thus shrink and disappear.   Equations used for critical radius in this simulation were taken from ref. \cite{Boyanovsky:2006bf}. 

The derivation for the wall velocity was found in \cite{Moore:1995si}. Discrete values of wall velocities for a corresponding Higgs mass from \cite{Moore:1995si} were interpolated in Mathematica in order to find a value for a Higgs mass of 35 GeV.  The wall velocity was approximated to be $V_w=0.375$ c, where $c$ is the speed of light.  

\section{Probability Functions}
Now that the critical radius and wall velocity have been described, the probability of bubble formation at a particular time, and the initial radius of a bubble are also needed for the simulation.

\subsection{Determining Initial Radius}
To determine the initial size of a bubble, a generic approximation of bubble nucleation was applied to the Electroweak Phase Transition. The probability function, $P_r$, is defined in \cite{Bravina:1995} and applied to the simulation.

When the function is plotted, the area under the curve represents the probability that a bubble will be formed within the corresponding domain. We note that the function is also temperature dependent so the distribution of bubbles will not be uniform with time.  A higher probability is given for a bubble to shrink near the start of the phase transition. However, as the temperature cools the function gives equal probabilities for radii sizes.

\subsection{Determining Formation Time}

The electroweak theory determines the probability that a bubble will form per unit time and per unit volume.  As the universe evolves in time, it also cools, so time is related to temperature. An approximation for the probability of formation with temperature, $P_t$, was found in \cite{Enqvist:1991xw} and applied to the simulation.

This probability function works the same way as $P_r$ in the way of determining the probability of a particular domain.  By examining Fig. \ref{ProbTimePlot}, it can be seen that bubbles are more likely to form near the end of the phase transition. 
\begin{figure}[ht] 
\centering
\includegraphics[width=0.8\linewidth]{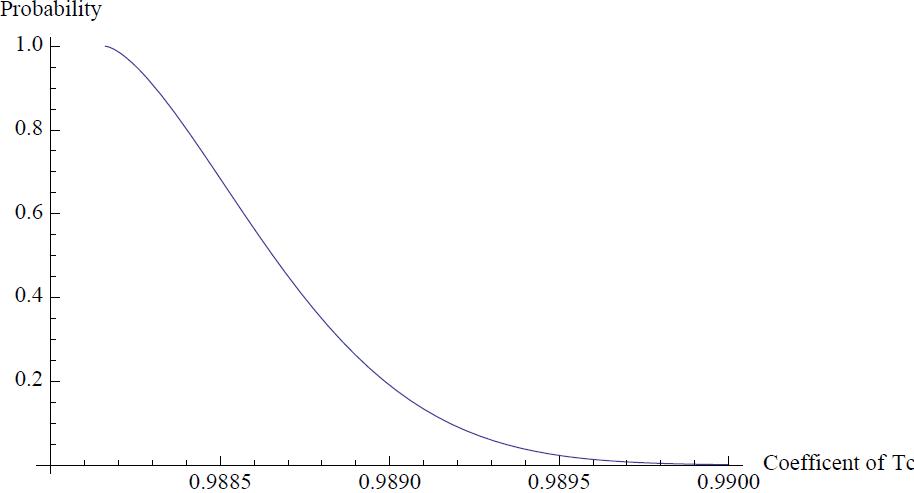} 
\caption{The probability of bubble formation vs. temperature.  Values before 0.9882 $T_c$ were complex.}
\label{ProbTimePlot}
\end{figure}

\section{Programing the Simulation}

The goal of this simulation was to demonstrate how these bubbles might form and grow over a certain area of space and time. 

\subsection{Scaling Space and Time}

In order to have a physical simulation the correct scaling for space and time had to be determined. For scaling time and temperature, the domain of the real values for $P_t$ was used. These ranged from 0.99 $T_c$ - 0.9882 $T_c$. This was then converted to a 10 second time scale. 
 
A way to establish the spacing of the bubbles was also needed for the simulation. From \cite{Enqvist:1991xw}, the equation $r_s$ will give the average spacing between a bubble and the nearest bubble. When thirty bubbles were programmed to form within a 300 x 300 unit in 2 dimensional space, the average spacing was found to converge to 29.7 units. 

\subsection{Generating Parameters}
\label{GeneratingParameters}
The next step was to program how to determine the parameters for a given bubble. The X and Y coordinates for the center of each bubble were randomly chosen with no weights. This proved to be the simplest method and since thirty bubbles were used, an even distribution often followed. The formation time of a given bubble was weighted by $P_t$ and the initial radius by $P_r$.  Next, an algorithm to reject bubbles that would form overlapping with other bubbles was written. 

After all of the parameters were calculated, they were exported to be simulated. The parameter set next was manipulated to the necessary format for sonification and then exported in the form of .csv files for the sonification process.

\section{Sonification}

A selection of 15 simulations have been made available on the video-sharing site ``YouTube,'' and can be found under the keyword `bubble phase transition.'  Screenshots from the 9th simulation are displayed in Fig. \ref{fig:bubbles}

 \begin{figure}[ht]
\centering
\begin{minipage}[b]{0.2\columnwidth}
  \centering
\includegraphics[height=1.5cm]{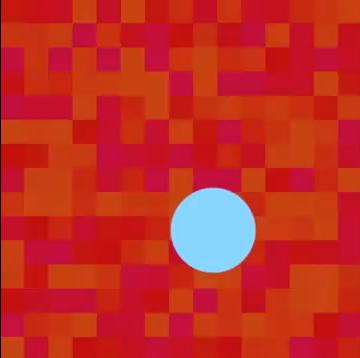} 
  \centerline{$t\approx4$(s)}
\end{minipage}
\begin{minipage}[b]{0.2\columnwidth}
  \centering
\includegraphics[height=1.5cm]{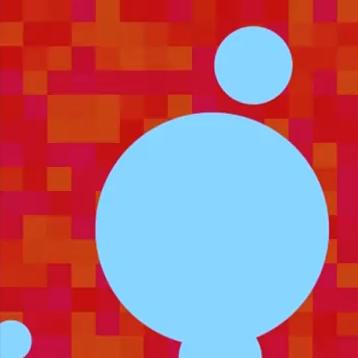} 
  \centerline{$t\approx7(s)$}
\end{minipage}
\begin{minipage}[b]{0.2\columnwidth}
  \centering
\includegraphics[height=1.5cm]{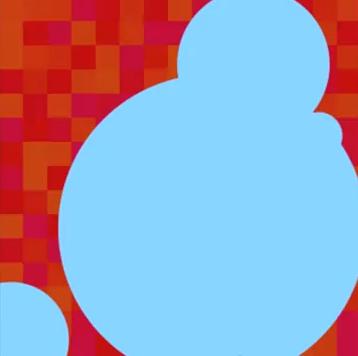} 
  \centerline{$t\approx8(s)$}
\end{minipage}
\begin{minipage}[b]{0.2\columnwidth}
  \centering
\includegraphics[height=1.5cm]{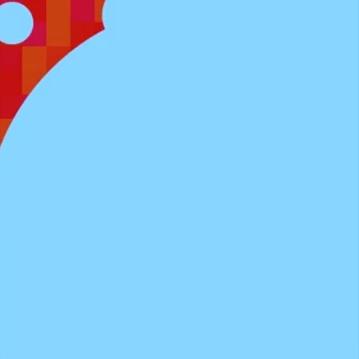} 
  \centerline{$t\approx9(s)$}
\end{minipage}
\caption{Four screenshots from the 9th simulation.  Before $t\approx4$ seconds, there has been no bubble nucleation.  After $t\approx10$ seconds, the phase transition is complete.}
\label{fig:bubbles}
\end{figure}

In order to enhance the simulation, a sonification was created using the program SuperCollider 3\cite{SC}.   Two synthesis definitions were employed, one to represent the formation of the bubbles over time, the other to represent the amount of space that still had Electroweak Symmetry.  The sonification techniques employed were judged to be effective based upon brief discussions of auditory perception with Dr. Neuhoff and Dr. Walker in 2009, various papers from the ICAD proceedings, and the current level of experience with SuperCollider.  However, insights from the international community would be beneficial to ongoing research goals.

\subsection{Sonification of Bubble Formation}

\label{SonificationA}

The sonification of bubble formation was modeled upon three known quantities, 
\begin{eqnarray}\label{csvA}\nonumber
&&\text{A) the time that the bubble first appeared,} \\\nonumber
&&\text{B) its horizontal (X) position, and} \\
&&\text{C) its vertical (Y) position,} 
\end{eqnarray}  
as generated using the previously described Mathematica program.  Mathematica wrote these values into a .csv file which was imported into SuperCollider.  Using A from (\ref{csvA}), two additional values were derived,
\begin{eqnarray}\label{csvB}\nonumber
&&\text{D) time until the end of the simulation, and} \\
&&\text{E) time between bubble events.}
\end{eqnarray}  
An example data set imported into SuperCollider is displayed in Table \ref{CSVTable}.

\begin{table}[h] 
	\centering
	\caption{An example data set imported into SuperCollider for a 13 second simulation.  A, B, C, D and E refer to the titles in (\ref{csvA}) and (\ref{csvB}). The values for A, D, E are in seconds and C, D are the relative X and Y positions on a 300 x 300 visual grid.}
	\label{CSVTable}
		\begin{tabular}{cccccc} 
Bubble \# & A  & B &C &D &E \\
\hline
1 &5.6 & 23.44&73.45& 7.4&	0.53\\
2 &6.13 & 254.97 & 91.67  &	6.87&	1.79	\\
3 &7.92 & 193.55 & 40.59  &	5.08 &	0.08	\\
4 &8 & 172.75 & 151.04 &  5 &	0.06	\\
5 &8.06 & 89.96 & 183.79 &	4.94&	0.71\\
6 &8.77 &161.62 & 65.08 & 4.23	&0	
		\end{tabular}
\end{table}

	As discussed in previous sections, each bubble that formed had an initial radius $r$ that would shrink or grow depending upon its magnitude. Values for $r$ that were below some critical radius $r_c$ would briefly appear and then disappear, whereas bubbles with $r>r_c$  lasted until the end of the simulation.  The bubbles which quickly shrank were not sonified because of lack of time and because a sufficient technique for sonification was not determined.  However, because instances of shrinking bubbles ($r<r_c$) were perceptually insignificant compared to the bubbles which expanded ($r>r_c$), our choice was not thought to detract from the overall simulation.

\subsubsection{Spatialization, Pitch, Timbre}
	To minimize perceptual interference between parameters and maximize aesthetic appeal, Spatialization, Timbre, Intensity and Pitch were the auditory parameters chosen for representation.  
	
	For every bubble with $r>r_c$ there was one corresponding sound event such that the bubble's X-position corresponded to horizontal spatialization, its Y-position corresponded to pitch (high Y-position $\rightarrow$ high pitch), and the increasing radius corresponds to a timbre and intensity which changes in a way intended to simulate the feeling of sudden appearance followed by expansion.  
	
	In SuperCollider, spatialization was controlled using `Pan2', a two channel equal power panner function. Pitch was controlled using `Midicps', a command which converts MIDI notes to their corresponding frequency.  The MIDI range used in the present sonification was 40-60 or alternatively, E2 to C4 in musical notation.  Timbre was controlled using `Formant,' a command which generates a set of harmonics around a formant frequency given a certain fundamental frequency.  Both the formant frequency and the pulse width frequency increased from 100-1000 Hertz along an exponential curve (`XLine') in the time frame of D-1 seconds. D-1 refers to the first element of the D column in Table \ref{CSVTable}.  Only the first bubble reaches the 1000 Hz level.

	So that the sonification would outlast the Electroweak Phase Transition, an extra three seconds were added to the simulation.  This additional time created a fuller auditory experience.  Because the data was not generated for a 13 second visualization, only visualizations in which the entire visual grid had undergone transition after 10 seconds were deemed physical.  The majority of the visualizations did not complete the phase transition within 10 seconds, consistent with \cite{Enqvist:1991xw}.  
	
\subsubsection{Intensity}
	The intensity of a bubble forming was also varied over time such that the increasing bubble size would have an analogous auditory experience.  In SuperCollider, the Envelope Generator (`EnvGen') was used to control the intensity over time.  An example envelope is displayed in Fig. \ref{envelope}.  The envelope was designed such that every bubble could be heard when it first entered, but by the end of the simulation, would not be as loud as the bubble which began before it.  The intensity was made to increase or decrease cubically to reflect the fact that the volume of a bubble is proportional to $r^3$.

\begin{figure}[ht] 
	\centering
	\includegraphics[width=0.8\linewidth]{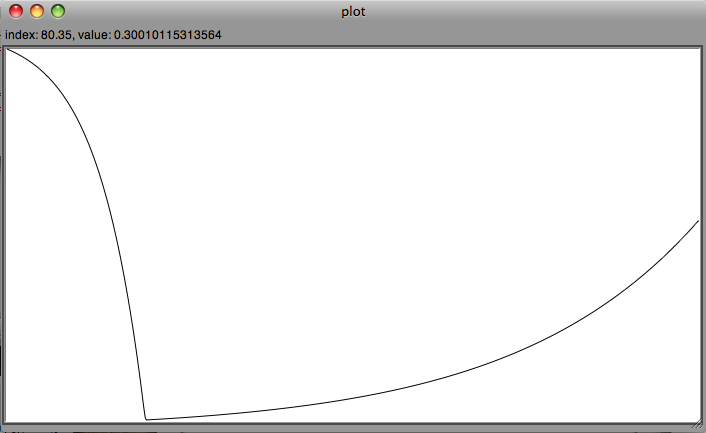} 
	\caption{An example envelope for controlling the intensity over time of each bubble.  This envelope corresponds to Bubble \#4 of Table \ref{CSVTable}.  The initial intensity of 1 decreases to 0.3 cubically in the 1 time frame of second.  Having reached its minimum value it approaches its final intensity of 5/7.4 cubically in the time frame of 5-1 seconds. Note that the entire envelope occurs over the time frame of D-4 of Table \ref{CSVTable}.}
	\label{envelope}
\end{figure}
	
	While designing the sonification, it was determined that bubbles with lower pitch (lower Y-coordinate), sounded softer than higher pitch bubbles (higher Y-coordinate).  Though the exact relationship for perceived equal intensity across octaves was not known, an additional intensity function was created to compensate for this inequality.  The function was constructed so that the amplitude of the lowest possible pitch (E2) was twice that of the highest possible pitch (C4), with a linear relationship between the endpoints. 
	
\subsection{Sonification of Electroweak Symmetry}

If the sonification involved only bubble formation, about the first five seconds of the simulation would be completely silent and featureless, somewhat at odds with the reality of a hot universe streaming with particles moving at the speed of light.  To represent the fluctuations in energy present in this phase of the universe, a crackling noise was chosen for the area which had not yet undergone transition.  In SuperCollider, this crackling noise was generated using `Dust2', a function which generates a certain density of random impulses from -1 to 1.  The average number (density) of these random impulses per second was chosen to be a constant 1000.  

As bubbles are formed, the area which still has has electroweak symmetry slowly decreases.  For this period of time, the intensity of the dust decreases proportional to the amount of space that still has electroweak symmetry.  As in section \ref{SonificationA}, the Envelope Genererator `EnvGen' was used for its control of intensity.  An example envelope can be viewed in Fig. \ref{Dust}.

 \begin{figure}[ht] 
	\centering
	\includegraphics[width=0.8\linewidth]{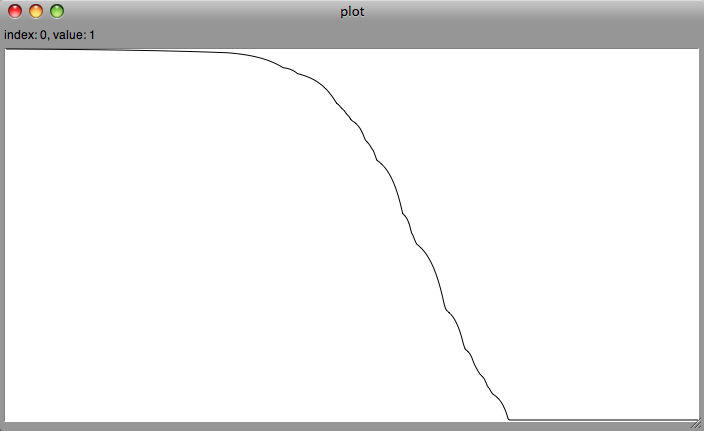} 
	\caption{A screenshot from SuperCollider displaying the envelope of `Dust2' over a period of 13 seconds.  Beginning with an initial intensity of 1, as electroweak symmetry is broken the envelope decreases to a final value of 0 when electroweak symmetry is completely broken.}
	\label{Dust}
\end{figure}

The envelope was created using the arrays generated in section \ref{GeneratingParameters} that determined whether bubbles would form inside other bubbles.  Though more rigorous calculations could be employed to calculate this area over time, they were not attempted because of lack of time.  Furthermore, the perceivable difference in sound between the two methods was not thought to be substantial enough to merit further inquiry.

\subsection{Analysis of Sonification}

The goal of the sonification was to augment a scientific visualization for a fuller and more engaging perceptual experience. The most important data properties were the bubble's horizontal and vertical position, the amount of unbroken electroweak symmetry remaining, and the rate of expansion for each bubble.  The horizontal and vertical position of the bubble centers, indicated by the spatial positioning and pitch respectively, communicates the areas of space that have undergone baryogenesis in each simulation.  This data property is complemented by the volume envelope, providing an additional cue as to the total amount of remaining electroweak symmetry.  The perceptual feeling of expansion (increasing volume) communicates the dynamic characteristics of the bubble wall.

The sonification created to model electroweak symmetry was thought to be effective because it communicated the important data properties in an intuitive way.  Its effectiveness was confirmed when errors in the Mathematica code were discovered through instances in which the combined visualization and sonification were not of the same data set. In one instance, a bubble was heard forming inside of another bubble (an unphysical event), which was not evident from the visual display.  This discrepancy indicated an error in the Mathematica code that may otherwise have gone undetected.

Several issues involving the intensity of the sound files arose.  It was discovered that lower pitches were softer than higher pitches despite being equal in all other respects.  The SuperCollider UGen `AmpComp' provides basic psychoacoustic amplitude compensation and would have helped to provide more uniformity.  Further progress towards loudness scaling for digital synthesis has been realized in \cite{Tipei:2004}.

Though the envelope of Fig. \ref{envelope} gave special emphasis to bubbles that had just formed, at some points it proved insufficient in this respect.  There were instances in which new bubbles were not heard as clearly as others because of the intensity of the combined sounds of other bubbles which had previously formed.  Though perhaps these new events are less relevant when there are many bubbles expanding simultaneously, an envelope in Fig. \ref{envelope} that started from 0 with a shorter attack time would have increased the perception of new bubbles.

Physically speaking, at the phase transition, the universe does not cool drastically.  While the sonification effectively demonstrated the phase transition, the intensity of the dust sound is not as great as the intensity of the sounds after the transition.  This discrepancy might be misleading and could be altered to keep a constant intensity over time.  However, this adjustment might have lead to a decrease in perception of bubble formation and ultimately was not used.

Another option for displaying the amount of electroweak symmetry over time would be altering the density of `Dust2' over time instead of its intensity.  Although this possibility was considered, it would have required the ability to do curve fitting in SuperCollider, a feature it does not support at this time.

\section{Conclusions}

The sonification presented was effective as a complement to visualization and as a tool for experiencing the dynamics of a phase transition that occurred shortly after the Big Bang. The overlap of sonification and visualization offered a means of finding and correcting bugs that could not have been discovered through visualization alone.  The sonification created would benefit from insights and corrections from the science of auditory perception.  Simulations such as these have potential usefulness in public outreach and physics education.

\section{Future Work}

The actual process of baryogenesis occured at a rate 10 trillion times faster than that displayed in our simulation.  If this simulation were slower or faster by a power of 10, humans likely find it less aesthetically pleasing and more difficult to interpret.  However, much finer adjustments can be made to this simulation.  Is there an ideal rate that sonic events in this simulation (or any simulation) would occur and evolve?

A broader area of research in the intersection between science and aesthetics is 1/f noise and stochastic processes in nature generally.  Given that the electroweak phase transition is stochastic, happening at random points in space governed by a temperature-dependent probability, it has a certain natural appeal for listeners.  Following the discovery of 1/f noise in music and speech\cite{Voss:1978}, algorithms have been developed which can predict the ``pleasantness" of musical and non-musical works\cite{Manaris:2005}.   Might their algorithm have some place in a program such as SuperCollider?

Although SuperCollider is a powerful tool for sound synthesis, as a physics research project involving complex computations, Mathematica was the computer program of choice.  Mathematica is common and well-established tool for computations but is clearly limited in its capacity for sound production.  Likewise, there are many powerful and well-established computer programs in a variety of academic and non-academic sectors in which capacity for sound production, though likely beneficial, is limited or non-existent.  The scope and speed of the evolution of sonification and auditory display would improve with the integration of sound production into these programs.

\section{Acknowledgments}

The authors would like to thank Michael Dine for his thoughtful advice and suggestions for this paper.  This research was supported by the NSF-REU grant DMR-0649112 and the College of Wooster.

\end{sloppy}
\end{document}